\documentclass[]{raa}           

\usepackage{graphicx,times}
\usepackage{natbib}
\usepackage{amssymb,amsmath}
\usepackage{longtable}
\bibpunct{(}{)}{;}{a}{}{,}

\usepackage[pagebackref=true]{hyperref}

\begin{document}

   \title{Long-term speckle interferometric monitoring of binary systems II: 2007-2025 positional measurements and improvement of orbits}

   \setcounter{page}{1}

   \author{Arina Mitrofanova \inst{}, Vladimir Dyachenko \inst{}, Anatoly Beskakotov \inst{}, Alexander Maksimov \inst{}, Margarita Butorina \inst{}, 
   	Yuri Balega \inst{}, Denis Rastegaev \inst{}}

   \institute{ Special Astrophysical Observatory, 369167 Nizhnij Arkhyz, Russia; {\it mitrofanova@sao.ru}}
\vs \no

\abstract{At the end of the last century, several research groups worldwide started monitoring binary and multiple systems with low-mass components, motivated by the results of the Hipparcos mission. Its goal was to construct precise orbits and calculate fundamental parameters, and therefore to refine empirical relations (such as the mass–luminosity relation). Despite the results achieved in this area, the orbital solutions of some objects require improvement. This article is focused on the analysis of observational data and improvement of orbital solutions of 8 objects (HIP~1055, HIP~2532, HIP~15633, HIP~19472, HIP~20227, HIP~20751, HIP~21710 and HIP~101181). The listed objects are members of a sample of approximately 300 nearby (d $<$ 100 pc) multiple systems compiled for long-term monitoring at the 6-meter telescope of the Special Astrophysical Observatory of the Russian Academy of Sciences (BTA SAO RAS). Speckle interferometric observations were carried out from 2007 to the present, which corresponds to 19 years of monitoring of the systems. Therefore, the new measurements are more numerous than or comparable to those in the literature, which makes it possible to make the observation series more complete. A long observation series made it possible to cover from ~20\% to ~70\% of the phases of the orbital periods of the objects under study. In combination with previously published data, the percentage of orbital coverage by measurements ranged from ~45\% to ~85\%. As a result, 4 orbits were classified as "definitive" (Grade~1), one as "good" (Grade~2), 2 as "reliable" (Grade~3), and one as "preliminary" (Grade~4) according to a grading scheme suggested by W.I. Hartkopf, B.D. Mason and C.E. Worley. Despite the high grades of the orbital solutions, it is recommended to continue monitoring (in particular, to obtain periastron observations) for HIP~1055, HIP~2532, HIP~15633, HIP~19472 and HIP~20751 in order to derive the final orbital parameters. It is not possible to obtain an accurate orbital solution for HIP~20227 due to measurements whose assigned weight strongly influences the period value, and for HIP~21710(Aa,Ab) due to the long orbital period ($P_{orb}$ $>$ 200 yr) and, consequently, the lack of observational data. An analysis and comparison of the mass sums and masses of components obtained by two independent methods and based on several parallax values were also carried out. In most cases, the best agreement is found between fundamental parameters calculated using Gaia parallaxes, if available.
\keywords{techniques: high angular resolution --- stars: binaries: visual --- stars: fundamental parameters --- stars: low mass --- stars: solar type --- stars: individual: HIP~1055, HIP~2532, HIP~15633, HIP~19472, HIP~20227, HIP~20751, HIP~21710, HIP~101181}
}

   \authorrunning{A. Mitrofanova et al. }            
   \titlerunning{Long-term speckle interferometric monitoring of binary systems II}  
   \maketitle

%
\section{Introduction}           
\label{sect:intro}

Most stars are members of binary and multiple systems. Binaries are a reliable source for determining stellar masses empirically. Knowledge of precise orbital parameters and the component masses underpins our understanding of a wide range of phenomena, from the initial mass function to the evolution of close binaries and progenitors of supernovae. To derive the accurate masses, precise parallaxes and complete and accurate solutions of the orbital equations are required. In turn, high-angular-resolution observations covering a significant part of the orbital period are required to obtain orbital parameters with high accuracy. Alongside space missions (for example, Gaia mission \citep{Gaia_16_novA1}), speckle interferometry at ground-based telescopes remains highly effective for these purposes.  

Speckle interferometry was invented by \citet{Labeyrie_70_may} and provides diffraction-limited accuracy of positional parameters of the objects under study. Along with research teams around the world (for example, \citet{Davidson_24_mar, Kalari_25_nov, Mason_23_oct, Mendez_25_apr, Tokovinin_24_jul}), the group of high-resolution methods in astronomy at the Special Astrophysical Observatory of the Russian Academy of Sciences (SAO RAS) continues long-term monitoring of binaries at the 6 m telescope. The diffraction limit for the Big Telescope Alt-azimuth (BTA) is 0.02 arcseconds at a wavelength of 500 nm, which allows measurements to be taken at epochs close to periastron passage, as well as to study short-period systems \citep{Mitrofanova_21_oct}. However, modern research is based not only on a large number of positional measurements but also on the parallaxes from the Gaia mission, which are of great importance \citep{Gaia_18_aug, Gaia_23_jun}.

This is the second in a series of publications \citep{Mitrofanova_24_sep}, in which we continue to publish the results of long-term monitoring of nearby (d $<$ 100 pc) binary and multiple systems with low-mass components. This sample of approximately 300 objects was compiled using stellar multiplicity data obtained by the Hipparcos mission \citep{ESA_97_jan}. Although this sample was compiled a considerable time previously, the number of positional measurements for its objects remained insufficient to construct their precise orbits. The solution to this problem was systematic monitoring of the objects at the 6-meter BTA telescope by our team \citep{Mitrofanova_20_jun, Mitrofanova_21_oct, Mitrofanova_24_sep} and by a number of teams worldwide (series of works by \citet{Hartkopf_15_oct} at the USNO 26 inch telescope, \citet{Horch_17_may} at 3.5 m WIYN telescope, \citet{Tokovinin_24_jul} at 4.1 m SOAR telescope, \citet{Clark_22_aug} at 4.3 m Lowell Discovery and 3.5 m WIYN telescopes, among others), which made it possible to construct orbital solutions for some of the systems, evaluate their quality, calculate the mass sums and component masses, and verify the accuracy of the existing parallaxes. Also, monitoring binary and multiple systems by several research teams is particularly advantageous given the high workload of large telescopes with the required diffraction limit and, accordingly, the limited amount of observing time that may not be carried out due to weather conditions.

\begin{figure} 
	\centering
	\includegraphics[width=7.2cm, angle=0]{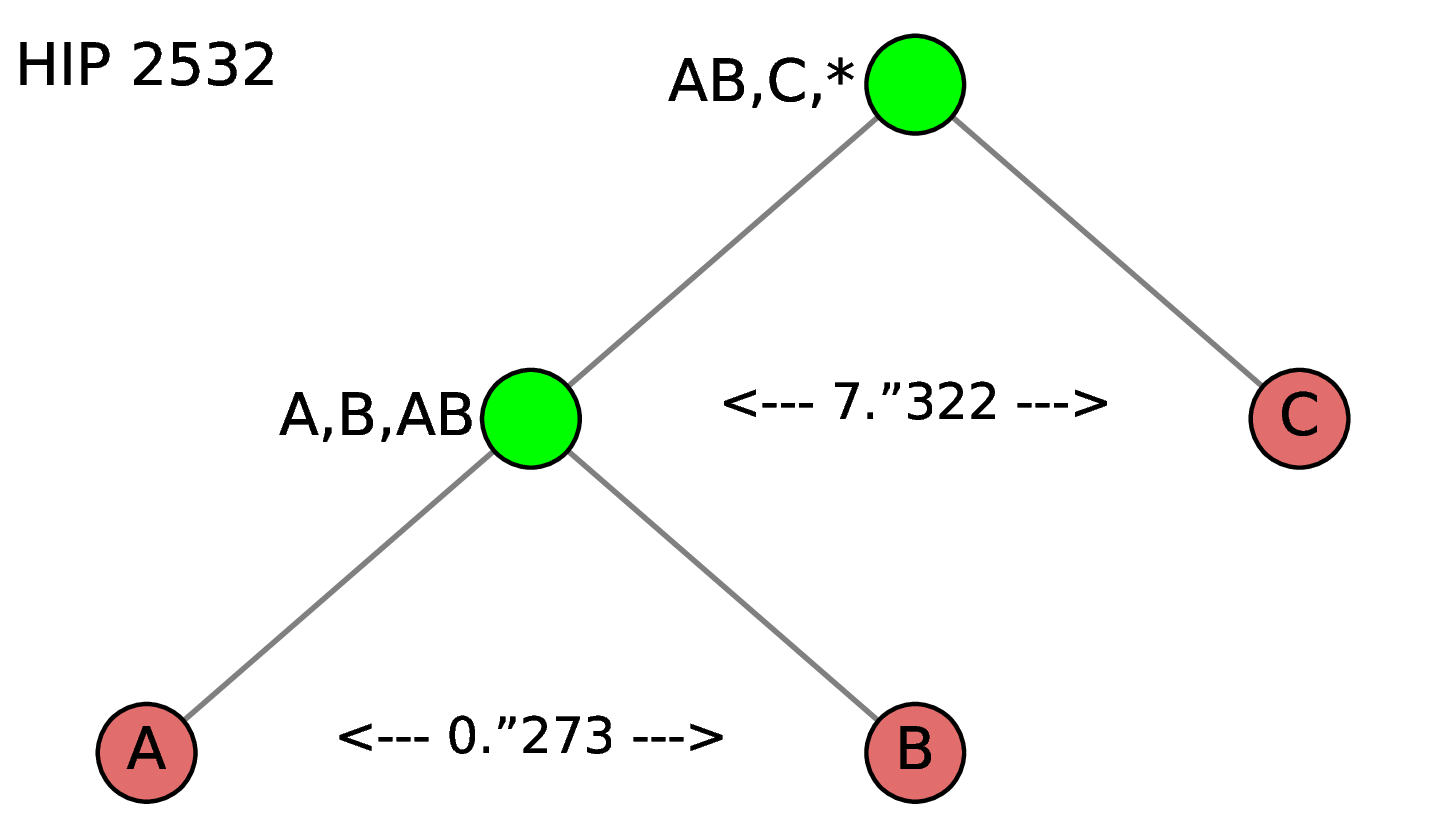}
	\includegraphics[width=7.2cm, angle=0]{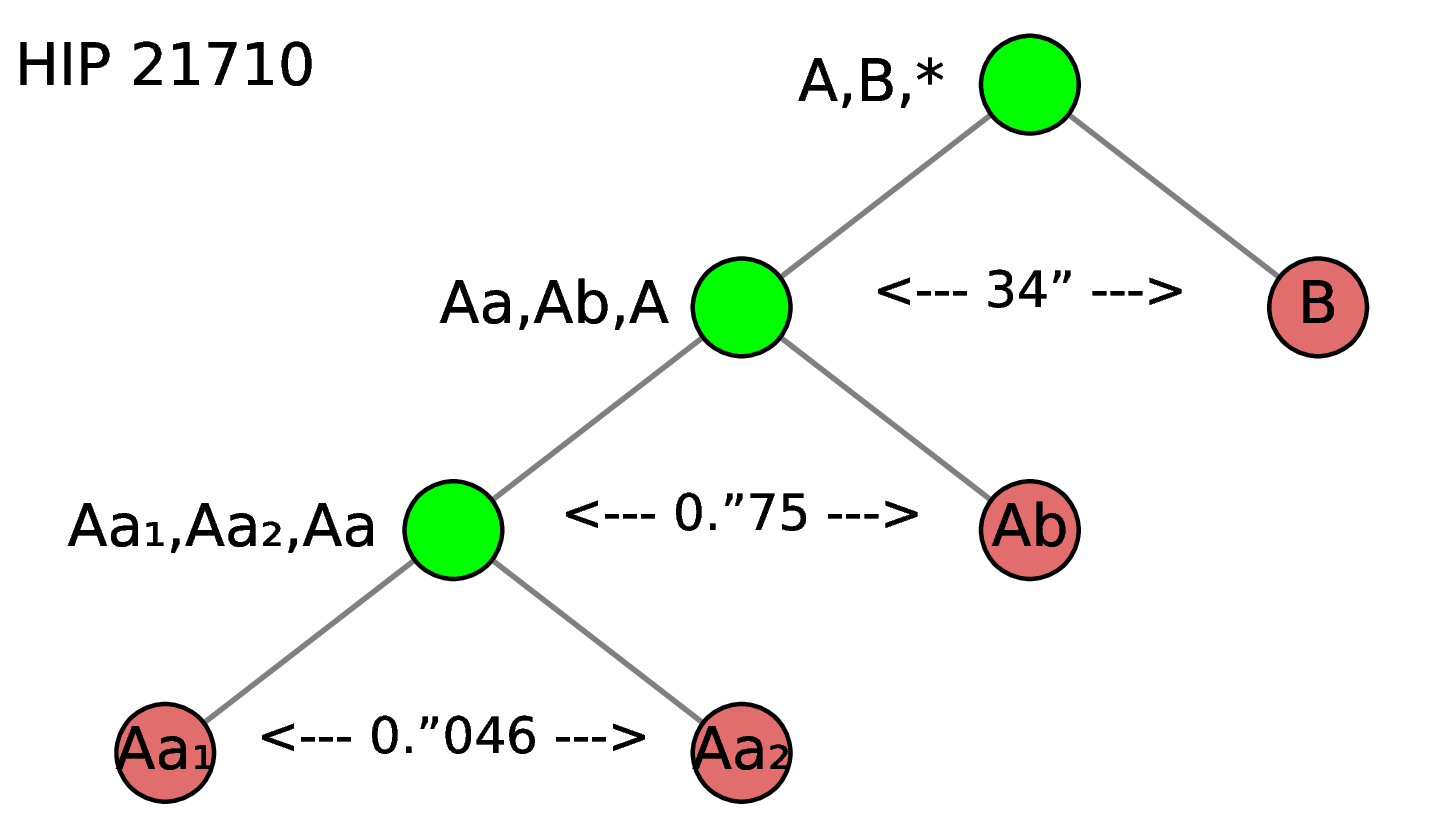}
	\caption{Schematic representation of the triple HIP~2532 and quadrupole HIP~21710 hierarchies indicating the separation between the components.} 
	\label{Fig1}
\end{figure}

In this article, we present the results of 19 years of speckle interferometry monitoring of 8 more systems from the above sample: HIP~1055, HIP~2532, HIP~15633, HIP~19472, HIP~20227, HIP~20751, HIP~21710 and HIP~101181. The preliminary values of the periods of these systems exceed 20 years, and therefore their monitoring over such a long time (for HIP~20751 almost equal to the value of the orbital period) is justified, and for some of the systems under study it will allow measurements to cover a large part of the phases of the orbital period. In this sample, 6 systems are binaries with preliminary semi-major axes of $\sim$115-420 mas, HIP~2532 is triple \citep{Tokovinin_23_apr} and HIP~21710 is quadruple \citep{Beuzit_04_oct}. Figure \ref{Fig1} schematically shows the hierarchical architecture of HIP~2532 and HIP~21710, and also indicates the separation between components (for HIP~2532 from \citet{Masda_25_feb}; for HIP~21710 from \citet{Beuzit_04_oct, Sperauskas_19_jun, Tokovinin_21_mar}).

We report new measurements, improve orbital elements, and calculate fundamental parameters for objects with well-determined parallaxes. The paper is structured as follows: Section 2 describes the observational data and data reduction results. Section 3 presents the improved orbits, their classification and fundamental parameters calculated using the new orbital parameters. The results of the study are discussed in Section 4.

\section{Observations. Data reduction}
\label{sect:Obs}

Speckle interferometric monitoring of target objects was carried out from 2007 to 2025 at the BTA of the SAO RAS utilizing a speckle interferometer \citep{Maksimov_09_jul}. During the observations, three EMCCD detectors were used to obtain images: PhotonMAX-512B (until 2010), Andor iXon+ X-3974 (2010-2014), and Andor iXon Ultra 897 (since 2015). A standard exposure time of 20 milliseconds and series of 1940 (until 2010) and 2000 images were used to record speckle interferograms. The following interference filters were used (central wavelength $\lambda$ / bandpass $\Delta \lambda$): 550/20 nm, 550/50 nm, 600/40 nm, 700/50 nm, 800/100 nm, and 900/80 nm. Most of the speckle images ($\sim$63\%) were obtained under good weather conditions with seeing of 1\arcsec-2\arcsec, $\sim$10\% of the observations were carried out under excellent weather conditions with seeing $<$1\arcsec, $\sim$21\% of the data were obtained under satisfactory weather conditions with seeing 2\arcsec-2.5\arcsec, and $\sim$6\% of the data were carried out under bad weather conditions with seeing $>$2.5\arcsec. 

High measurement accuracy is ensured by the simultaneous use of several calibration methods applied to speckle interferometric images \citep{Mitrofanova_20_jun}. Positional parameters and magnitude differences were determined through the analysis of the power spectrum and the autocorrelation function of the speckle interferometric series, as described in \cite{Balega_02_apr} and \cite{Pluzhnik_05_feb}. The reconstruction of the position of the secondary was carried out using bispectral analysis \citep{Lohmann_83_dec}.  

Table~\ref{tab1} presents the observation log, as well as new and previously published positional parameters and the magnitude differences of the systems under study. The columns are object designation; epoch of observation in fractions of Besselian year; telescope diameter; bandpass or $\lambda$/$\Delta \lambda$; $\theta$ and $\sigma_{\theta}$ are the position angle of the secondary relative to the primary and its error; $\rho$ and $\sigma_{\rho}$ are the separation between the two stars and its error; $\Delta m$ and $\sigma_{\Delta m}$ are the magnitude difference and its error; and reference. The analysis of the magnitude differences from the data obtained in different epochs shows that the population standard deviation of new measurements is about 0.09 mag. 

\begin{table}
	\bc
	\begin{minipage}[]{100mm}
		\caption[]{Positional Parameters and the Magnitude Differences.\label{tab1}}\end{minipage}
	\setlength{\tabcolsep}{1pt}
	\small
	\begin{tabular}{ccccccccccc}
		\hline\noalign{\smallskip}
Object & Epoch & Telescope & $\lambda$/$\Delta\lambda$ & $\theta\degr$ & $\sigma_{\theta}\degr$ & $\rho$ & $\sigma_{\rho}$ & $\Delta m$ & $\sigma_{\Delta m}$ & Reference \\
& yr & diameter, m & nm &  & & mas & mas & mag & mag &  \\
		\hline\noalign{\smallskip}
HIP~1055 & 1991.25 & 0.29 &  & 181.0 &  & 357 &  &  &  & \citet{ESA_97_jan} \\
& 1998.7718 & 6 & 800/60 & 173.5 & 0.50 & 558 & 2 & 1.13 & 0.03 & \citet{Balega_02_apr} \\
& 1999.7470 & 6 & 2115/214 & 172.3 & 1.00 & 565 & 3 & 0.84 & 0.12 & \citet{Balega_02_apr} \\
& 1999.8156 & 6 & 800/60 & 172.4 & 0.2 & 578.7 & 0.7 & 1.15 & 0.03 & \citet{Balega_04_aug} \\
& 2000.8755 & 6 & 800/110 & 172.2 & 0.2 & 594 & 3 & 1.09 & 0.05 & \citet{Balega_06_jan} \\
  \noalign{\smallskip}\hline
\end{tabular}
\ec
\tablecomments{0.86\textwidth}{Table~\ref{tab1} is available in machine-readable form and lists the parameters published earlier by \cite{Balega_02_apr, Balega_04_aug, Balega_06_jan, Balega_07_dec, Balega_13_jan, Beuzit_04_oct, Couteau_90_may, ESA_97_jan, Heintz_80_oct, Heintz_87_sep, Heintz_90_sep, Heintz_95_aug, Horch_02_jun, Horch_04_mar, Horch_08_jul, Horch_09_jun, Horch_10_jan, Horch_11_feb, Horch_11_jun, Horch_12_jan, Horch_17_may, Horch_21_jun, Mason_93_jan, Mason_99_apr, Orlov_09_oct, Patience_98_may, Tokovinin_10_feb, Tokovinin_15_aug, Tokovinin_16_jun, Tokovinin_18_jun, Tokovinin_19_jul, Tokovinin_20_jul, Tokovinin_21_aug, Tokovinin_22_aug, Tokovinin_24_jul}.}
\end{table}

\section{Orbit construction}
\label{sect:Orbits}

The orbital parameters of the objects under study were calculated sequentially. First, a preliminary orbital solution was constructed using the method of Monet \citep{Monet_77_jun}. All measurements presented in Table \ref{tab1} were used to construct the orbital solution. During the construction of the preliminary orbit, the quadrant ambiguities were identified in the published positional measurements. It should be noted that quadrant ambiguities were most often responsible for the erroneous construction of orbital solutions (especially for objects with large orbital periods) when long series of measurements had not yet been accumulated (e.g., HIP~111546 \citep{Mitrofanova_21_oct}, Chara~122Aa \citep{Mitrofanova_24_sep}). With modern data series that accurately determine the position of the secondary, identifying quadrant ambiguities is straightforward. As a result, quadrant ambiguities were identified for the positional parameters of HIP~2532, HIP~19472, HIP~20751 and HIP~101181, and the values of these position angles were changed by 180\degr. Previously, when constructing the orbits of these systems, quadrant ambiguities were corrected (moreover, the number of such measurements is only 1-2 per system), so the orbital solutions were improved without significant changes. However, we decided to list these measurements additionally in the object descriptions to present a complete picture. Table~\ref{tab1} shows the values of the position angles in accordance with the published data, and those changed by 180\degr~are listed in a separate descriptions below for each system. 

Then, to construct the final orbit, we used the ORBIT software package \citep{Tokovinin_92}. This stage typically requires 2-3 iterations. Since each measurement obtained with different instruments and methods contributes to the orbital solution and has its systematic error, we assigned corresponding weights to them. The weight assigned to a measurement depends on the values of the residuals and deviations from the orbit (smaller weights are assigned to those with large errors). First, we evaluate the measurements that have the largest residuals relative to the preliminary orbital solution. Examples of such measurements are 1991.25 \citet{ESA_97_jan} for HIP~2532, 1987.91 and 1989.84 \citet{Heintz_90_sep}, 1989.72 \citet{Couteau_90_may} and 1993.99 \citet{Heintz_95_aug} for HIP~15633, 1979.00 \citet{Heintz_80_oct}, 1986.99 \citet{Heintz_87_sep} and 2004.1126 \citet{Horch_08_jul} for HIP~19472, and 1996.0217 \citet{Patience_98_may} for HIP~20751. Most often, such measurements are not taken into account when constructing the final orbit, or they are assigned minimal weights so that their contribution becomes insignificant. In our case, the listed measurements were excluded from the analysis. After this, orbital solutions tend to be more stable during fitting. Next, several orbits are usually constructed as follows: (1) all remaining measurements are given the same maximum weight; (2) measurements with the largest residuals that remain after excluding “bad” data are given smaller weights. As a result, orbits are selected that better fit the observational data, with smaller $\chi^{2}$ and the absence of systematic deviations in the residuals for $\rho$ and $\theta$. However, for some systems (in this case HIP~20227, and in the case of GJ~3076 from \citet{Mitrofanova_24_sep}) measurements (e.g., those from \citet{ESA_97_jan}) are crucial for constructing the orbit, but experience shows that these data often have larger residuals than others. In such cases, further monitoring of the object is undoubtedly necessary. Therefore, orbital residuals in $\rho$ and $\theta$ can be higher than the estimates of uncertainties presented in Table~\ref{tab1}.

The orbital solution of the triple system HIP~21710(Aa,Ab) was constructed with the IDL code \mbox{orbit3.pro} \citep{Tokovinin_17_mar, Tokovinin_17_feb_orb3}. This code is used when observational data are available for both the inner and outer pairs (radial velocities and astrometric data), and allows these measurements to be fitted simultaneously. In the case of HIP~21710(Aa,Ab) there are radial velocities for the inner $Aa_{1},Aa_{2}$ pair and positional measurements for the outer Aa,Ab pair. Simultaneous fitting of two orbits allows for the motion relative to the center of mass of the internal system to be taken into account, for which a special wobble factor is introduced. The wobble factor for HIP~21710(Aa,Ab) is positive (more details below in the system description) because the $Aa_{1},Aa_{2}$ pair belongs to the primary. A negative wobble factor would indicate configuration Aa,($Ab_{1},Ab_{2}$). This program also allows one to fix the elements of the inner and outer orbits, and they do not change later during fitting.

Table~\ref{tab2} presents both our orbital parameters for the systems under study and those found in the literature. The columns give: (1) object designation; (2) the orbital period in years; (3) the epoch of periastron passage; (4) the eccentricity; (5) the semi-major axis in mas; (6) the longitude of the ascending node; (7) the argument of periastron; (8) the inclination; and (9) the reference for the calculation.

{\small\setlength{\tabcolsep}{0pt}
\begin{longtable}{ccccccccc}
		\caption{Orbital Parameters.}\label{tab2}\\
		\hline\noalign{\smallskip}
Object & $P_{orb}$, year & $T_{0}$, year & $e$ & $a$, mas & $\Omega$, $\degr$ &  $\omega$, $\degr$ & $i$, $\degr$ & Reference \\
		\hline\noalign{\smallskip}
HIP~1055 & 78.8 & 2062.4 & 0.895 & 423.3 & 331.7 & 348.3 & 136.4 & \citet{Horch_21_jun} \\
& $\pm 2.3$ & $\pm 2.1$ & $\pm 0.021$ & $\pm 6.7$ & $\pm 3.1$ & $\pm 4.0$ & $\pm 7.6$ & \\ \cline{2-9}
& 77.3 & 2057.9 & 0.756 & 435 & 341.5 & 358.7 & 116.1 & This work \\
& $\pm 5.9$ & $\pm 4.5$ & $\pm 0.023$ & $\pm 5$ & $\pm 1.4$ & $\pm 1.5$ & $\pm 1.2$ & \\
\hline
HIP~2532(AB) & 58.65 & 1993.26 & 0.754 & 355 & 27.7 & 276.3 & 124.6 & \citet{Cvetkovic_13_feb_circ} \\ \cline{2-9}
& 57.943 & 2034.432 & 0.233 & 266.6 & 126.6 & 325.3 & 132.3 & \citet{Cvetkovic_14_mar} \\
& $\pm 1.966$ & $\pm 0.950$ & $\pm 0.096$ & $\pm 3.8$ & $\pm 9.7$ & $\pm 21.0$ & $\pm 7.9$ & \\ \cline{2-9}
& 62.2 & 2035.84 & 0.216 & 273 & 124.9 & 320.2 & 131.5 & \citet{Tokovinin_24_feb_circ} \\ \cline{2-9}
& 40.796 & 2028.755 & 0.604 & 230 & 90.91 & 301.70 & 141.21 & \citet{Masda_25_feb} \\
& $\pm 0.49$ & $\pm 0.247$ & $\pm 0.006$ & $\pm 1$ & $\pm 0.51$ & $\pm 0.42$ & $\pm 0.19$ & \\ \cline{2-9}
& 40.1 & 2030.2 & 0.646 & 212 & 65.5 & 285.3 & 152.4 & This work \\
& $\pm 0.7$ & $\pm 0.6$ & $\pm 0.029$ & $\pm 8$ & $\pm 5.7$ & $\pm 5.7$ & $\pm 6.6$ & \\ 
\hline
HIP~15633 & 28.25 & 2001.42 & 0.793 & 155 & 58.7 & 185.8 & 117.5 & \citet{Zirm_12_feb_circ} \\ \cline{2-9}
& 32.41 & 2001.35 & 0.806 & 165 & 55.6 & 180.4 & 114.6 & \citet{Tokovinin_21_feb_circ} \\
& $\pm 0.64$ & $\pm 0.13$ & $\pm 0.012$ & $\pm 1$ & $\pm 1.3$ & $\pm 3.8$ & $\pm 1.9$ & \\ \cline{2-9}
& 32.0 & 2033.6 & 0.775 & 165 & 56.9 & 182.1 & 112.6 & This work \\
& $\pm 0.7$ & $\pm 0.6$ & $\pm 0.009$ & $\pm 1$ & $\pm 0.5$ & $\pm 0.5$ & $\pm 0.7$ & \\
\hline
HIP~19472 & 28.91 & 2000.53 & 0.832 & 282 & 165.9 & 35.9 & 115.7 & \citet{Rica_11_jun_circ} \\ \cline{2-9}
& 29.96 & 2000.513 & 0.833 & 287 & 165.1 & 34.1 & 116.9 & \citet{Tokovinin_21_feb_circ} \\
& $\pm 0.20$ & $\pm 0.017$ & $\pm 0.003$ & $\pm 2$ & $\pm 0.6$ & $\pm 1.2$ & $\pm 0.4$ & \\ \cline{2-9}
& 29.8 & 2030.3 & 0.838 & 278 & 166.1 & 32.2 & 119.2 & This work \\
& $\pm 0.7$ & $\pm 0.6$ & $\pm 0.015$ & $\pm 5$ & $\pm 0.7$ & $\pm 1.9$ & $\pm 2.2$ & \\
\hline
HIP~20227 & 73.483      & 2010.968    & 0.436       & 367       & 50.3       & 79.1       & 37.6      & \citet{Cvetkovic_17_dec} \\
& $\pm 0.911$ & $\pm 0.868$ & $\pm 0.081$ & $\pm 9.8$ & $\pm 11.7$ & $\pm 18.1$ & $\pm 6.0$  & \\ \cline{2-9}
& 59.9      & 2012.55    & 0.399       & 291     & 16.5     & 133.4     & 26.7      & This work \\
& $\pm 1.7$ & $\pm 0.09$ & $\pm 0.011$ & $\pm 6$ & $\pm 2.9$ & $\pm 3.0$ & $\pm 1.2$ & \\ \cline{2-8}
& 65.6      & 2012.77    & 0.437       & 313     & 13.3     & 139.3     & 30.0      &  \\
& $\pm 0.9$ & $\pm 0.03$ & $\pm 0.006$ & $\pm 2$ & $\pm 1.7$ & $\pm 1.6$ & $\pm 0.5$ & \\
\hline 
HIP~20751 & 19.77 & 2017.39 & 0.694 & 201 & 119.6 & 273.7 & 63.0 & \citet{Tokovinin_21_aug, Tokovinin_21_feb_circ} \\
& $\pm 0.27$ & $\pm 0.11$ & $\pm 0.040$ & $\pm 10$ & $\pm 1.5$ & $\pm 0.9$ & $\pm 2.2$  & \\ \cline{2-9}
& 19.2 & 2036.4 & 0.586 & 182 & 122.5 & 272.0 & 60.5 & This work \\
& $\pm 0.1$ & $\pm 0.1$ & $\pm 0.008$ & $\pm 2$ & $\pm 0.7$ & $\pm 0.4$ & $\pm 0.5$ & \\
\hline
HIP~21710 & 204.0 & 2028.17 & 0.124 & 1302 & 154.1 & 133.5 & 77.1 & \citet{Tokovinin_17_oct_circ} \\
&  &  &  &  &  &  &  & (Aa,Ab) \\ \cline{2-9}
& 610.02 (d) & 47474.27 (d) & 0.399 &  &  & 266.96 &  & \citet{Halbwachs_18_nov} \\
& $\pm 1.02$ & $\pm 5.96$ & $\pm 0.023$ &  &  & $\pm 4.75$ &  & ($Aa_{1},Aa_{2}$) \\ \cline{2-9}
& 610.43 (d) & 56635.1 (d) & 0.415 &  &  & 269.9 &  & \citet{Sperauskas_19_jun} \\
& $\pm 0.15$ & $\pm 4.2$ & $\pm 0.018$ &  &  & $\pm 3.4$ &  & ($Aa_{1},Aa_{2}$) \\ \cline{2-9}
& 1.6732 & 2013.958 & 0.426 & 46 & 286.5 & 267.7 & 62.7 & \citet{Tokovinin_21_mar} \\
& $\pm 0.0005$ & $\pm 0.013$ & $\pm 0.018$ & fixed & $\pm 10.2$ & $\pm 3.6$ & $\pm 14.3$ & ($Aa_{1},Aa_{2}$) \\ 
& 285 & 1999.8 & 0.190 & 1637 & 335.7 & 253.0 & 78.6 &  \\
& fix & $\pm 76$ & $\pm 0.090$ & $\pm 136$ & $\pm 4.9$ & $\pm 145$ & $\pm 1.5$ & (Aa,Ab) \\ \cline{2-9}
& 1.6732 & 2013.958 & 0.426 & 46 & 286.5 & 267.7 & 62.7 & This work \\
& fixed & fixed & fixed & fixed & fixed & fixed & fixed & ($Aa_{1},Aa_{2}$) \\
& 285 & 2030.8 & 0.197 & 1637 & 333.9 & 311.64 & 79.16 &  \\
& fixed & $\pm 0.3$ & $\pm 0.006$ & fixed & $\pm 0.1$ & $\pm 1.07$ & $\pm 0.08$ & (Aa,Ab) \\ \cline{2-8}
& 1.6704 & 2013.873 & 0.414 & 41 & 282.5 & 259.4 & 75.8 &  \\
& $\pm 0.0005$ & $\pm 0.009$ & $\pm 0.016$ & 2 & $\pm 2.4$ & $\pm 2.8$ & $\pm 2.8$ & ($Aa_{1},Aa_{2}$) \\ 
& 202.1 & 2017.0 & 0.086 & 1263 & 154.2 & 108.07 & 77.2 &  \\
& fixed & $\pm 0.4$ & $\pm 0.006$ & fixed & $\pm 0.1$ & $\pm 1.07$ & $\pm 0.2$ & (Aa,Ab) \\ 
\hline
HIP~101181 & 26.1 & 2006.12 & 0.829 & 122.4 & 170.5 & 309.5 & 66.9 & \citet{Horch_11_jun} \\
& $\pm 0.6$ & $\pm 0.09$ & $\pm 0.011$ & $\pm 3.9$ & $\pm 2.0$ & $\pm 2.5$ & $\pm 1.6$ & \\ \cline{2-9}
& 95.0 & 1991.3 & 0.392 & 254 & 152.2 & 337.3 & 82.0 & \citet{Tokovinin_19_oct_circ} \\ \cline{2-9}
& 31.65 & 2006.25 & 0.859 & 121 & 175.9 & 305.6 & 63.7 & \citet{Docobo_21_oct_circ} \\
& $\pm 1.3$ & $\pm 0.05$ & $\pm 0.003$ & $\pm 5$ & $\pm 1.0$ & $\pm 1.0$ & $\pm 1.5$ & \\ \cline{2-9}
& 48.64 & 1984.84 & 0.280 & 171 & 147.4 & 272.0 & 79.9 & \citet{Tokovinin_23_jun_circ} \\
& $\pm 0.93$ & $\pm 0.90$ &  & $\pm 2$ & $\pm 0.6$ & $\pm 3.1$ & $\pm 0.4$ & \\ \cline{2-9}
& 27.03 & 2011.67 & 0.577 & 114.9 & 137.2 & 34.3 & 70.5 & \citet{Tokovinin_24_oct_circ} \\
& $\pm 0.78$ & $\pm 0.41$ & $\pm 0.030$ & $\pm 3.3$ & $\pm 1.7$ & $\pm 6.5$ & $\pm 1.5$ & \\ \cline{2-9}
& 27.08 & 2038.97 & 0.572 & 114.2 & 138.09 & 35.7 & 71.8 & This work \\
& $\pm 0.03$ & $\pm 0.04$ & $\pm 0.001$ & $\pm 0.2$ & $\pm 0.07$ & $\pm 0.1$ & $\pm 0.1$ & \\
  \noalign{\smallskip}\hline
\end{longtable}
}

As in previous studies \citep{Mitrofanova_20_nov, Mitrofanova_20_jun, Mitrofanova_21_oct, Mitrofanova_24_sep, Efremova_21_apr}, we used two independent methods to determine the fundamental parameters of systems and their components (masses, spectral types and mass sum). The mass sum was calculated using the improved orbital parameters and Kepler's third law. Then it was compared with the component masses obtained from the data of  \citet{Pecaut_13_sep} based on absolute magnitudes of components calculated by applying Pogson's ratio \citep{Pogson_56_nov} using the average magnitude difference of the components in the 550 bandpass and the apparent magnitudes of the stars from the SIMBAD database. The following parallaxes were also used for calculations: Hipparcos \citep{van_Leeuwen_07_nov_plx}, Gaia DR2 and DR3 \citep{Gaia_18_aug, Gaia_23_jun}, and parallaxes previously published for individual systems. Table~\ref{tab3} lists the fundamental parameters of objects and their components. The columns give: (1) object designation; (2) the average magnitude difference of the components in the 550 bandpass ($\Delta m_{550}$); (3, 6) the absolute magnitudes of the components in V band ($M_ {A}$ and $M_ {B}$); (4, 7) their spectral types ($Sp_ {A}$ and $Sp_{B}$); (5, 8) the masses of stars ($\mathfrak{M}_{A}$ and $\mathfrak{M}_{B}$); (9) the mass sum of the components defined by orbital parameters by the first method ($\sum \mathfrak{M}$); (10) parallax source and its value; and (11) reference. Below, we show the orbits obtained and discuss each system in detail. 

{\small\setlength{\tabcolsep}{0pt}
	\begin{longtable}{ccccccccccc}
		\caption{Fundamental Parameters of the Objects.}\label{tab3}\\
		\hline\noalign{\smallskip}
		Object & $\Delta m_{550}$, & $M_{A}$, & $Sp_{A}$ & $\mathfrak{M}_{A}$, & $M_{B}$, & $Sp_{B}$ & $\mathfrak{M}_{B}$, & $\sum \mathfrak{M}$, & Parallax source & Reference \\
		& mag & mag &  & $\mathfrak{M}_{\odot}$ & mag &  & $\mathfrak{M}_{\odot}$ & $\mathfrak{M}_{\odot}$ & and value, mas & \\
\hline\noalign{\smallskip}
HIP~1055 & 1.89 & 7.52 & K5.5V & 0.66 & 9.41 & M0.5V- & 0.54- & 0.98 & Hipparcos & This work \\
& $\pm 0.09$ & $\pm 0.09$ &  &  & $\pm 0.13$ & M1V & 0.49 & $\pm 0.32$ & $24.15 \pm 2.28$ &  \\ \cline{3-10}
&  & 7.51 & K5.5V & 0.66 & 9.40 & M0.5V- & 0.54- & 0.99 & Gaia DR2 &  \\
&  & $\pm 0.09$ &  &  & $\pm 0.13$ & M1V & 0.49 & $\pm 0.16$ & $24.0695 \pm 0.1164$ &  \\ \cline{3-10}
&  & 7.51 & K5.5V & 0.66 & 9.40 & M0.5V- & 0.54- & 0.99 & Gaia DR3 &  \\
&  & $\pm 0.09$ &  &  & $\pm 0.13$ & M1V & 0.49 & $\pm 0.15$ & $24.0513 \pm 0.0281$ &  \\ \cline{3-10}
&  & 7.31 & K5V & 0.68 & 9.20 & M0.5V & 0.54 & 1.3 & \citet{VanAltena_95} &  \\
&  & $\pm 0.09$ &  &  & $\pm 0.13$ &  &  & $\pm 2.7$ & $21.9 \pm 15.0$ &  \\ 
\hline
HIP~2532(AB) & 0.44 & 4.63 & G2 & 1.11 & 5.07 & G5 & 1.05 & 1.86 & Hipparcos & \citet{Cvetkovic_14_mar} \\
& $\pm 0.16$ & $\pm 0.14$ &  &  & $\pm 0.21$ &  &  & $\pm 0.36$ & $14.47 \pm 0.84$ &  \\ \cline{2-11}
& 0.54 & 4.41 & F7.5V & 1.20 & 4.95 & G1.5V & 1.09 & 2.29 &  & \citet{Masda_25_feb} \\
& $\pm 0.02$ & $\pm 0.09$ &  & $\pm 0.07$ & $\pm 0.10$ &  & $\pm 0.06$ &  &  &  \\ \cline{2-11}
& 0.56 & 5.30 & G7V- & 0.96- & 5.86 & K0V- & 0.87- & 1.96 & Hipparcos & This work \\
& $\pm 0.11$ & $\pm 0.11$ & G8V & 0.94 & $\pm 0.16$ & K1.5V & 0.82 & $\pm 0.41$ & $14.47 \pm 0.84$ &  \\ \cline{3-10}
&  & 5.20 & G7V & 0.96 & 5.77 & K0V & 0.87 & 2.23 & \citet{Masda_25_feb} &  \\
&  & $\pm 0.11$ &  &  & $\pm 0.16$ &  &  & $\pm 0.25$ & $13.8528 \pm 0.20$ &  \\
\hline
HIP~15633 & 0.45 & 4.10 & F9V & 1.14 & 4.55 & G1V & 1.07 & 3.68 & Hipparcos & This work \\
& $\pm 0.12$ & $\pm 0.12$ &  &  & $\pm 0.17$ &  &  & $\pm 1.55$ & $10.60 \pm 1.48$ &  \\
\hline
HIP~19472 & 1.54 & 6.70 & K3.5V & 0.73 & 8.24 & K7V- & 0.63- & 1.26 & Hipparcos & This work \\
& $\pm 0.04$ & $\pm 0.06$ &  &  & $\pm 0.07$ & K8V & 0.59 & $\pm 0.35$ & $26.98 \pm 2.27$ &  \\ \cline{3-10}
&  & 6.66 & K3V- & 0.75- & 8.20 & K7V & 0.63 & 1.33 & Gaia DR3 &  \\ 
&  & $\pm 0.06$ & K3.5V & 0.73 & $\pm 0.07$ &  &  & $\pm 0.14$ & $26.4764 \pm 0.1088$ &  \\ 
\hline
HIP~20227 & 3.98 & 4.14 & F9 & 1.11 & 8.12 & K7 & 0.55 & $1.54 \pm 0.31$ & Hipparcos & \citet{Cvetkovic_17_dec} \\
& $\pm 0.07$ & $\pm 0.13$ &  &  & $\pm 0.15$ &  &  &  & $18.11 \pm 1.12$ & \\ \cline{2-11}
& 3.55 & 5.19 & G7V & 0.96 & 8.73 & K9V & 0.56 & 1.16 (1.20) & Hipparcos & This work \\
& $\pm 0.32$ & $\pm 0.32$ &  &  & $\pm 0.45$ &  &  & $\pm 0.24$ ($\pm 0.23$) & $18.11 \pm 1.12$ &  \\ \cline{3-10}
&  & 5.02 & G5V & 0.98 & 8.57 & K8V- & 0.59- & 1.45 (1.51) & Gaia DR2 &  \\
&  & $\pm 0.32$ &  &  & $\pm 0.45$ & K9V & 0.56 & $\pm 0.13$ ($\pm 0.06$) & $16.7836 \pm 0.1464$ &  \\ \cline{3-10}
&  & 5.06 & G5V- & 0.98- & 8.61 & K9V & 0.56 & 1.38 (1.43) & Gaia DR3 &  \\
&  & $\pm 0.32$ & G6V & 0.97 & $\pm 0.45$ &  &  & $\pm 0.12$ ($\pm 0.05$) & $17.0817 \pm 0.0946$ &  \\ \cline{3-10}
&  & 5.13 & G6V & 0.97 & 8.68 & K9V & 0.56 & 1.25 (1.29) & \citet{Cvetkovic_17_dec} &  \\
&  & $\pm 0.32$ &  &  & $\pm 0.45$ &  &  & $\pm 0.19$ ($\pm 0.17$) & $17.66 \pm 0.75$ &  \\ 
\hline
HIP~20751 & 1.63 & 6.56 & K3V & 0.75 & 8.19 & K7V & 0.63 & 1.17 & Hipparcos & This work \\
& $\pm 0.02$ & $\pm 0.02$ &  &  & $\pm 0.03$ &  &  & $\pm 0.23$ & $24.11 \pm 1.59$ &  \\ \cline{3-10}
&  & 6.29 & K2.5V & 0.76 & 7.92 & K6V- & 0.65- & 1.70 & Gaia DR2 &  \\
&  & $\pm 0.02$ & &  & $\pm 0.03$ & K6.5V & 0.64 & $\pm 0.12$ & $21.2806 \pm 0.4454$ &  \\ \cline{3-10}
&  & 6.28 & K2.5V & 0.76 & 7.91 & K6V- & 0.65- & 1.73 & Gaia DR3 &  \\
&  & $\pm 0.02$ &  &  & $\pm 0.03$ & K6.5V & 0.64 & $\pm 0.11$ & $21.1594 \pm 0.3952$ &  \\
\hline
HIP~21710(Aa,Ab) &  &  &  & 0.90 &  &  & 0.53 & 1.4 &  & \citet{Tokovinin_21_mar} \\ \cline{2-11}
& 3.49 &  &  &  &  &  &  & 1.41 & Hipparcos & This work \\
& $\pm 0.11$ &  &  &  &  &  &  & $\pm 0.20$ & $32.69 \pm 1.51$ &  \\ \cline{3-10}
&  &  &  &  &  &  &  & 0.99 & Gaia DR2 &  \\
&  &  &  &  &  &  &  & $\pm 0.03$ & $36.8039 \pm 0.3853$ &  \\ \cline{3-10}
&  &  &  &  &  &  &  & 1.37 & Gaia DR3 &  \\
&  &  &  &  &  &  &  & $\pm 0.07$ & $32.9986 \pm 0.5305$ &  \\
\hline
HIP~101181 & 0.87 & 3.62 & F6V & 1.25 & 4.49 & G1V & 1.07 & 1.54 & Hipparcos & This work \\
& $\pm 0.15$ & $\pm 0.15$ &  &  & $\pm 0.21$ &  &  & $\pm 0.37$ & $10.96 \pm 0.87$ &  \\ \cline{3-10}
&  & 3.16 & F3V- & 1.43- & 4.03 & F8V & 1.18 & 2.92 & Gaia DR2 &  \\
&  & $\pm 0.15$ & F4V & 1.39 & $\pm 0.21$ &  &  & $\pm 0.24$ & $8.8571 \pm 0.2388$ &  \\ \cline{3-10}
&  & 3.23 & F4V & 1.39 & 4.10 & F9V & 1.14 & 2.65 & Gaia DR3 &  \\
&  & $\pm 0.15$ &  &  & $\pm 0.21$ &  &  & $\pm 0.25$ & $9.1483 \pm 0.2843$ &  \\
\hline\noalign{\smallskip}
\end{longtable}
}

\subsection{Discussion of individual systems}

{\bf HIP~1055} ($00^{h} 13^{m} 09\fs2, +20\degr 22\arcmin 56\farcs8$; BD+1920) is a binary with K7 and M0 components ($M_{A}$=7.9 mag and $M_{B}$=9.1 mag \citep{Balega_02_apr}). The first visual orbital elements were calculated by \citet{Horch_21_jun}. 

\begin{figure} 
	\centering
	\includegraphics[width=7.2cm, angle=0]{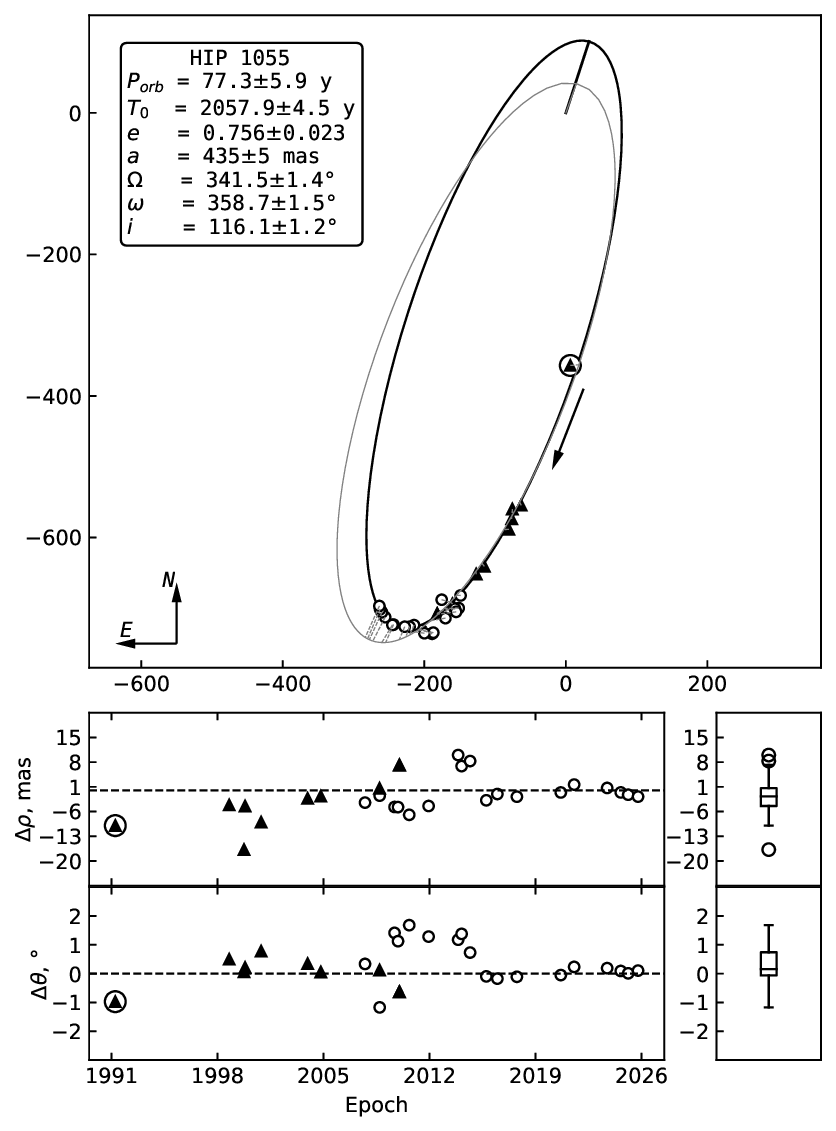}
	\includegraphics[width=7.2cm, angle=0]{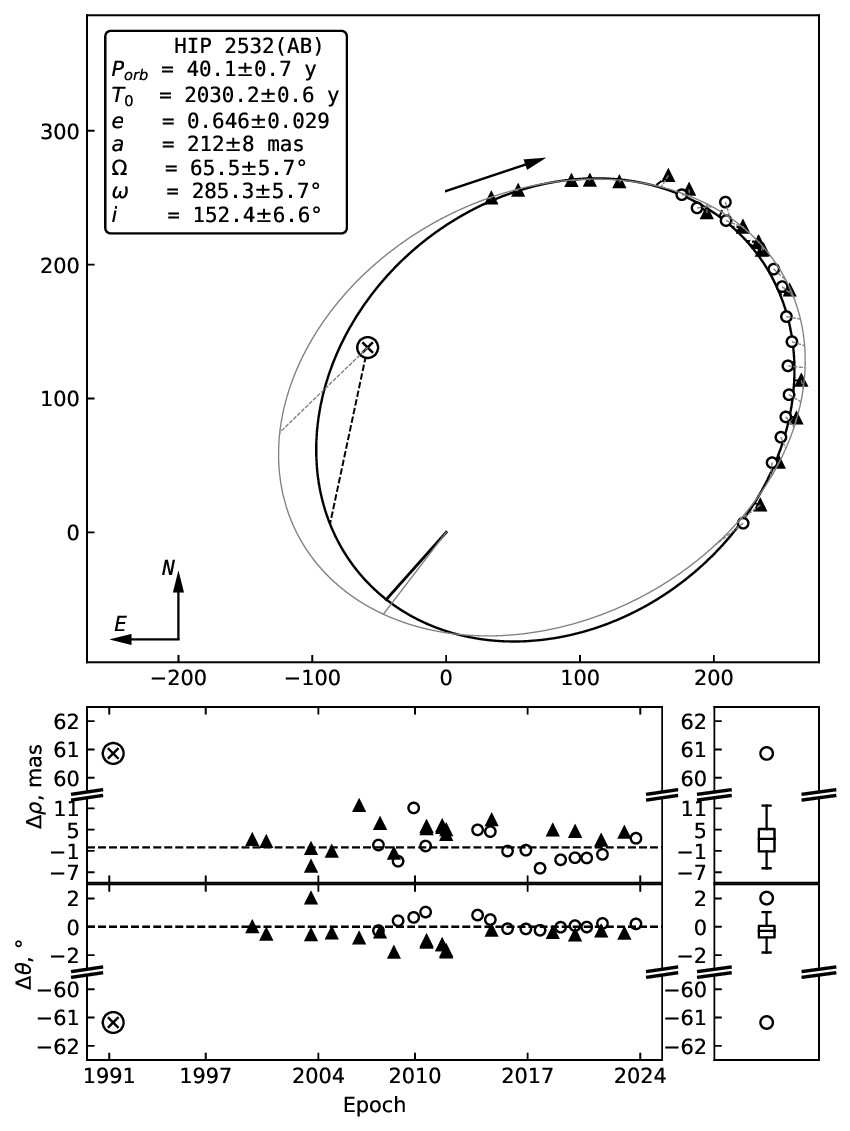}
	\includegraphics[width=7.2cm, angle=0]{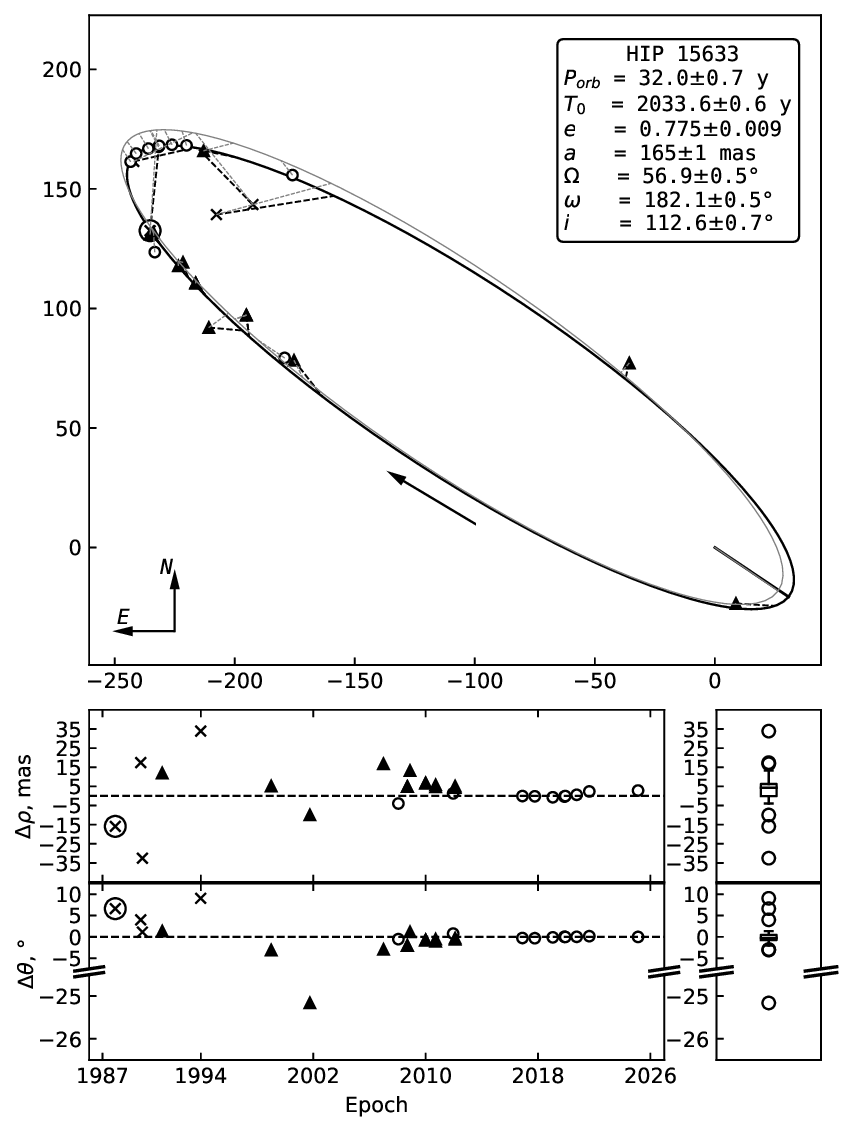}
	\includegraphics[width=7.2cm, angle=0]{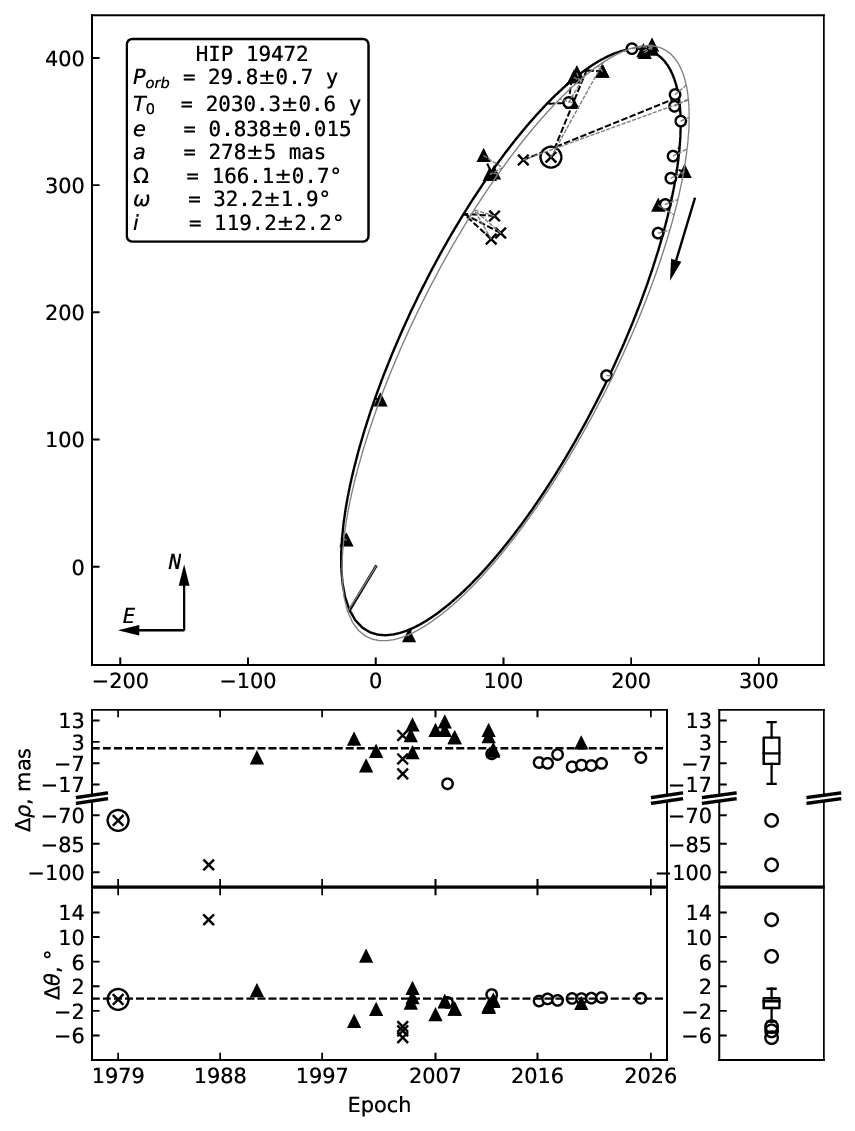}
	\caption{Orbital solutions for HIP~1055, HIP~2532(AB), HIP~15633 and HIP~19472. The orbits constructed in this work are marked in black. The previously published orbits are shown in gray. Triangles correspond to the published data; open circles - data obtained in this study; crosses - data with large residuals; a point placed in a large circle is the first measurement for system. The arrow shows the direction of motion of the secondary. $\Delta \rho$ and $\Delta \theta$ are residuals of angular separation and position angle showing difference between the observed and model value. The dashed line on the residuals plots indicates the orbital solution. Boxplots display the distribution of data based on a five-number summary (“minimum”, first quartile (Q1), median, third quartile (Q3), and “maximum”) and outliers.} 
	\label{Fig2}
\end{figure}

The orbital solution was improved by adding 18 new measurements to the 10 previously published (Figure \ref{Fig2}, top-left panel). Less than 50\% of the orbit is covered by measurements, and the residuals for $\rho$ and $\theta$ are 5.9 mas and 0\fdg7. Also, the values of the orbital period, epoch of periastron passage, and semi-major axis still have large uncertainties. Therefore, the orbit can only be classified as "reliable" (Grade 3), according to the classification of \citet{Hartkopf_01_dec}. The value $V_{J} = 10.43 \pm 0.01$ mag obtained by \citet{AlWardat_08_dec} was used to calculate the absolute magnitudes of the components. There are four parallaxes for this system (Table \ref{tab3}), three of which - from Hipparcos, Gaia DR2 and DR3 - have similar values. Consequently, the fundamental parameters differ little from each other and are in good agreement when calculated by the two methods. However, most of the measurements occur near apoastron, so observations of the object at periastron are necessary to confirm the orbital solution.

{\bf HIP~2532} ($00^{h} 32^{m} 07\fs1, -12\degr 17\arcmin 42\farcs3$; HD~2893) was detected as a binary by the Hipparcos mission. However, this system is triple, consisting of a binary AB and a component C at a separation of 7.332 arcseconds \citep{Tokovinin_23_apr, Masda_25_feb}. The study described in this article focuses on the primary, namely binary AB. The orbit for this system was initially constructed by \citet{Cvetkovic_13_feb_circ} and was further improved, which made it possible to calculate the total mass $\mathfrak{M}_{tot} = 1.86 \pm 0.36~\mathfrak{M}_{\odot}$ \citep{Cvetkovic_14_mar}. As a result of recently published spectrophotometric and dynamical analysis, \citet{Masda_25_feb} obtained improved orbital and fundamental parameters (Tables \ref{tab2} and \ref{tab3}) and a new value for the dynamical parallax $\pi_{dyn} = 13.8528 \pm 0.20$ mas.

The position angle values of the measurements by \citet{Balega_07_dec} and \citet{Horch_08_jul} (the only ones that correspond to the date 2006.53 yr) were changed by 180$\degr$ when constructing the orbit, which is shown in the top-right panel of Figure \ref{Fig2}. To improve the orbital solution, 34 measurements (20 previously published and 14 new) were used, which cover 80\% of the orbit. The residuals for $\rho$ and $\theta$ are 4.8 mas and 0\fdg8. The Hipparcos measurement has a large uncertainty and a gap of 8.6 years before the next measurement. Despite this, the remaining data allows for construction of a "good" orbit (60\% coverage, Grade 2), which can be slightly further improved by adding measurements near periastron. A parallax by Hipparcos and dynamical parallax by \citet{Masda_25_feb} are available for this object. The obtained mass sum $\sum \mathfrak{M} = 2.23 \pm 0.25 \mathfrak{M}_{\odot}$ using the parallax of \citet{Masda_25_feb} agrees well with the values from spectroscopy $\sum \mathfrak{M}_{sp} = 2.16 \mathfrak{M}_{\odot}$ \citep{Cvetkovic_14_mar} and spectrophotometry $\sum \mathfrak{M}_{Sph} = 2.29 \mathfrak{M}_{\odot}$ \citep{Masda_25_feb}.

{\bf HIP~15633} ($03^{h} 21^{m} 18\fs1, +10\degr 37\arcmin 47\farcs3$; HD~20779, HEI~499) is a binary, the first measurements of which were obtained by \citet{Heintz_90_sep} at the 0.6 m Lowell telescope.

The orbital solution for HIP~15633 was slightly improved after adding 10 new measurements to the 17 previously published ones. The residuals $\Delta \rho = 6.1$ mas and $\Delta \theta = 4\fdg9$ remain large despite the exclusion from the calculation of the measurements marked with crosses in the bottom-left panel of Figure \ref{Fig2}. This residual value is due to the use of all other measurements. Even though more than one orbital period has passed from the first to the last observation, the lack of data corresponding to the epoch of periastron passage does not allow for the construction of a "definitive" orbit (Grade 1). Therefore, it can be classified as "good" (Grade 2). As with the previous system, only the Hipparcos parallax is available for HIP~15633. Due to the large parallax error, the mass sum is calculated with a significant uncertainty of 40\%. Therefore, it can be said that the fundamental parameters agree well within the error limits. Improvement of the parallax value is clearly required for this object. 

{\bf HIP~19472} ($04^{h} 10^{m} 14\fs0, +17\degr 22\arcmin 06\farcs1$; HD~285465, HEI~35) was resolved as a binary by \citet{Heintz_80_oct} for the first time. The position angle of the measurement by \citet{Balega_06_jan} (2001.758) was changed by 180$\degr$ when constructing the improved orbit, which is shown in the bottom-right panel of Figure \ref{Fig2}. More than one period has passed since the beginning of observations of the object, and the data (22 previously published measurements and 10 new ones) cover almost all phases of the orbital period, with a maximum phase gap of 0.15. The orbital solution of HIP~19472 is classified as "definitive" (Grade 1). Data near periastron passage were obtained from 1999 to 2001, so new observations near this epoch are required for a final classification. To obtain the apparent magnitudes of HIP~19472 in the V band ($M_{V} = 9.31 \pm 0.05$ mag) we used photometric ratios and conversion factor from \citet{Busso_18_jul}. Fundamental parameters were calculated using the Hipparcos and Gaia parallaxes, which have similar values. The mass estimates obtained by the two methods are in good agreement.

{\bf HIP~20227} ($04^{h} 20^{m} 06\fs3, +31\degr 08\arcmin 26\farcs4$; HD~27323, HDS~555) is a binary, the first orbit of which was constructed by \citet{Cvetkovic_17_dec}. The orbital solution was based on six measurements, which made it possible to calculate a dynamical parallax of $17.66 \pm 0.75$ mas, and mass sums of $1.54 \pm 0.31~\mathfrak{M}_{\odot}$ and $2.10 \pm 0.30~\mathfrak{M}_{\odot}$ using the dynamical and Gaia DR1 ($16.33 \pm 0.62$ mas) parallaxes, respectively.

\begin{figure} 
	\centering
	\includegraphics[width=4.7cm, angle=0]{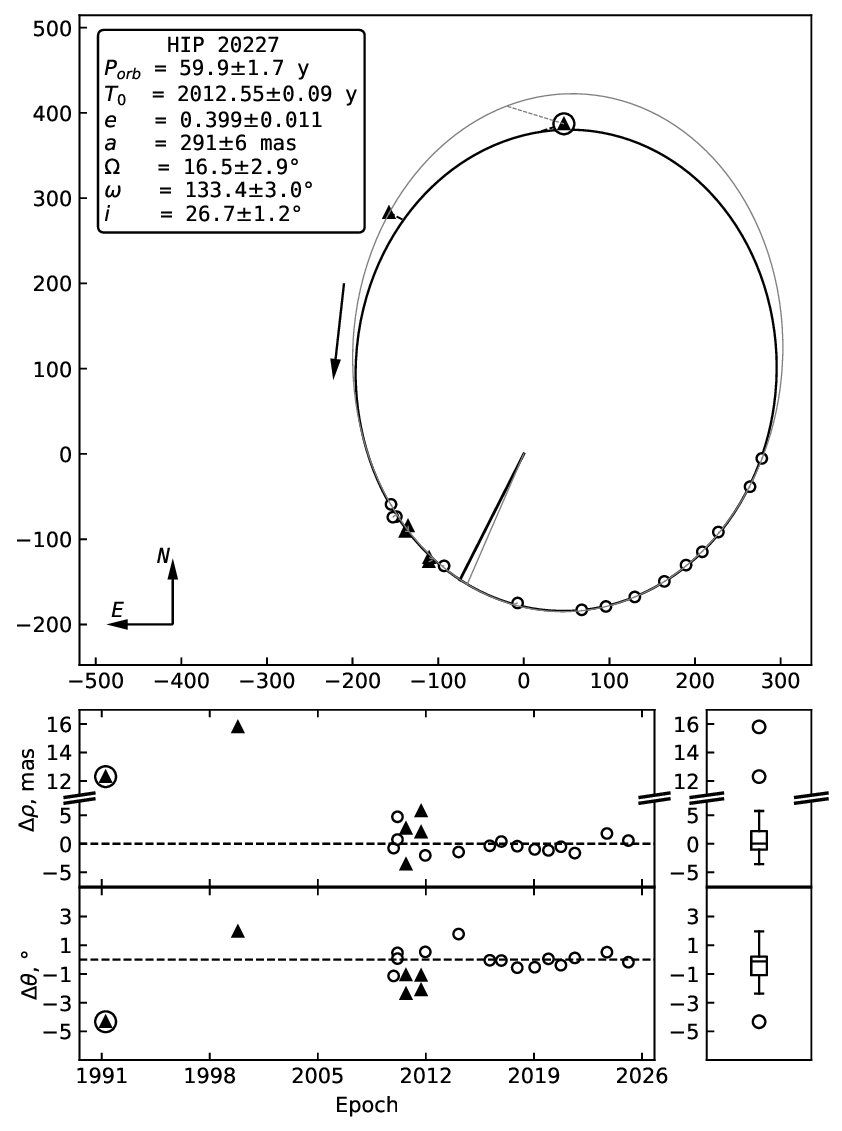}
	\includegraphics[width=4.7cm, angle=0]{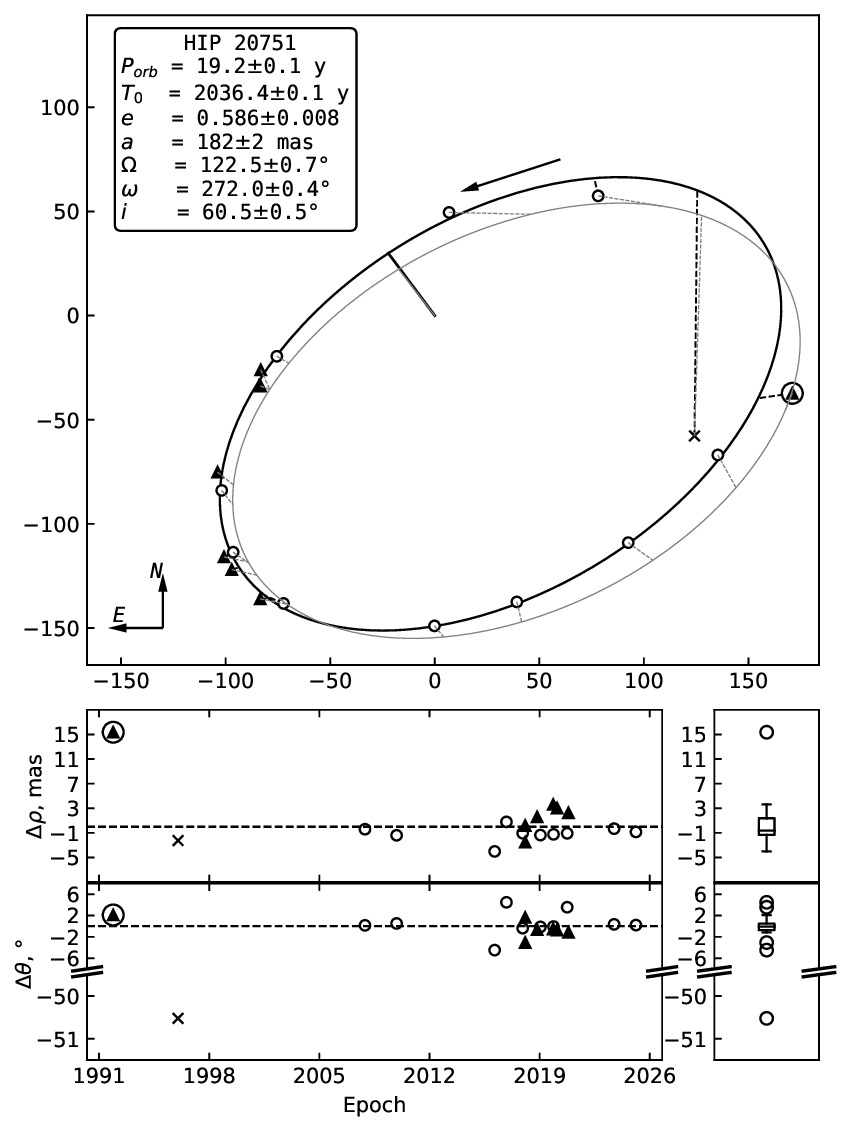}
	\includegraphics[width=4.7cm, angle=0]{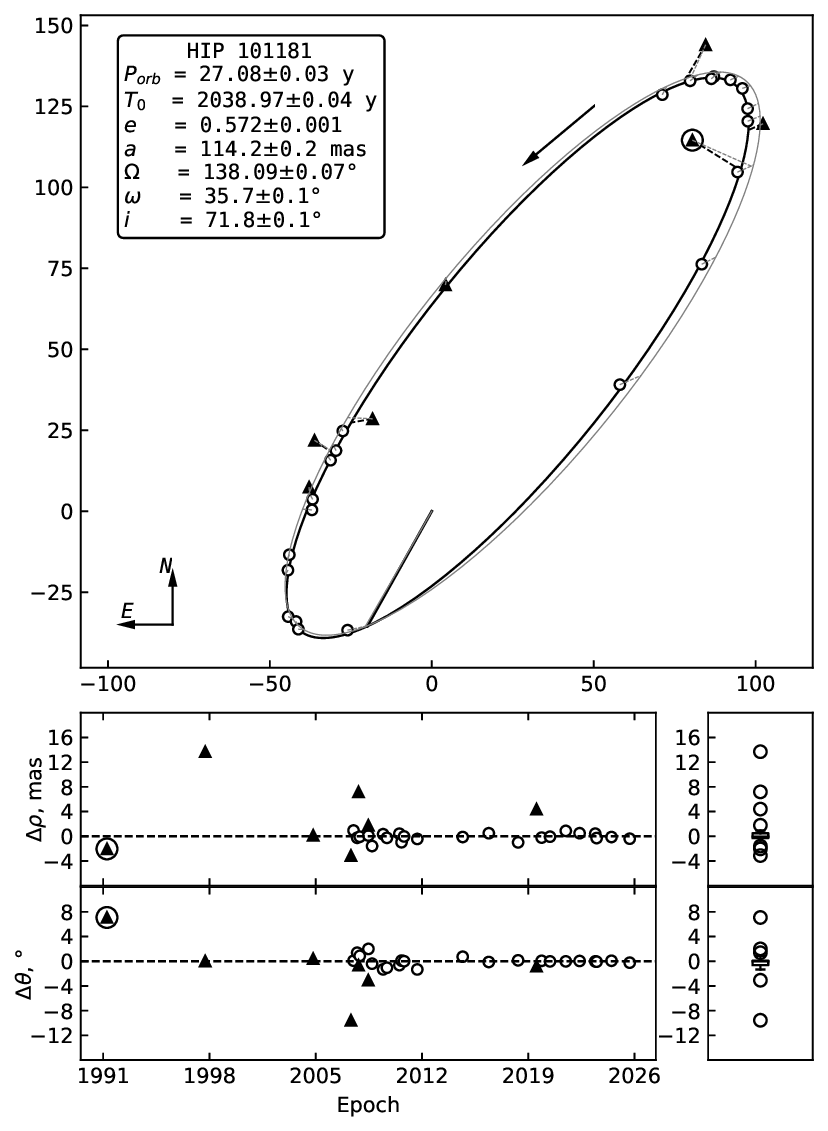}
	\caption{Orbital solutions for HIP~20227, HIP~20751 and HIP~101181. All designations are described in the caption of Figure \ref{Fig2}.} 
	\label{Fig3}
\end{figure}

As a result of long-term monitoring of HIP~20227, 14 new measurements were obtained in addition to the 6 previously published ones. The new orbital solution is very dependent on the weights of the measurements by Hipparcos and \citet{Balega_04_aug}. The two constructed orbits, black with a period of 59.9 years and gray with a period of 65.6 years, are shown in the left panel of Figure \ref{Fig3} and in Table \ref{tab2}. The periastron has already passed, and the slow movement will now begin, but further monitoring is needed (with a large interval between observations). The maximum gap in phase is 0.45 and a little more than half of the period has been covered since the system was first observed. Therefore, the orbit can be classified as "reliable" (Grade 3). To obtain the apparent magnitudes of HIP~20227 in the V band ($M_{V} = 8.86 \pm 0.05$ mag), we used photometric ratios and the conversion factor from \citet{Busso_18_jul}. The mass sums of the components calculated for an orbit with a period of 65.6 years are given in parentheses in Table \ref{tab3}. There are four parallaxes in total for this system: the Hipparcos parallax, the Gaia DR3 and DR2 parallaxes, and the parallax from \citet{Cvetkovic_17_dec}. As a result, the parameters obtained using Gaia DR2 parallax are in better agreement for both orbital solutions.

{\bf HIP~20751} ($04^{h} 26^{m} 48\fs3, +10\degr 52\arcmin 15\farcs8$; HD~286820, PAT~8) was resolved as a binary by \citet{Patience_98_may}. The first orbit with a period of 20 years and periastron in 2017.4 was constructed by \citet{Tokovinin_21_aug} based on observations at SOAR and on historical measurements.

The position angle values of the measurements by \citet{Mason_93_jan} and \citet{Patience_98_may} were changed by 180$\degr$ when constructing the improved orbit using 18 measurements (8 previously published and 10 new). The new orbital solution (Figure \ref{Fig3}, middle panel) is characterized by well-distributed coverage exceeding one period, and no revisions, except for minor adjustments, are expected. Therefore, the orbit is classified as "definitive" (Grade 1). The fundamental parameters of the object were calculated using the parallaxes from Hipparcos, Gaia DR2 and DR3. However, the masses and mass sums calculated for all three parallax values are not in good agreement. This means that further monitoring or parallax improvement is necessary.

{\bf HIP~21710} ($04^{h} 39^{m} 42\fs6, +09\degr 52\arcmin 19\farcs5$; HD~286955, HDS~601) is a quadruple system with periods of 23 kyr for the A,B pair, 285 yr for the Aa,Ab pair, and 610.7 days for the inner spectroscopic binary \citep{Beuzit_04_oct, Halbwachs_03_jan, Tokovinin_17_oct_circ, Halbwachs_18_nov, Sperauskas_19_jun, Tokovinin_21_mar}. The hierarchy of this system is schematically shown in the right panel of Figure \ref{Fig1}. \citet{Tokovinin_21_mar} constructed an orbital solution for the Aa,Ab pair based on speckle interferometric measurements (for the Aa,Ab pair) and radial velocities (for the $Aa_{1}$,$Aa_{2}$ pair), but noted large error values for some orbital elements. 

\begin{figure} 
	\centering
	\includegraphics[width=7.2cm, angle=0]{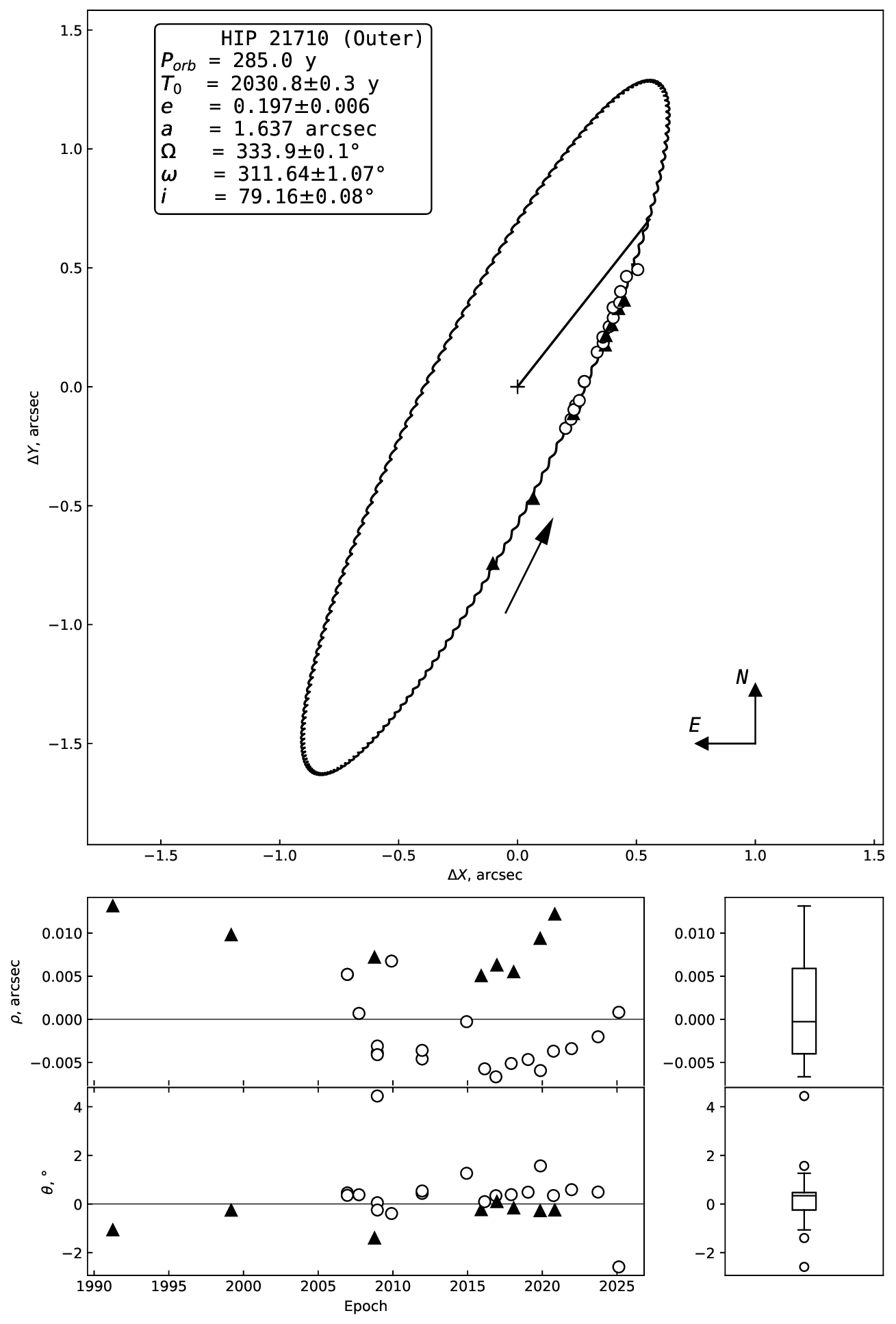}
	\includegraphics[width=7.2cm, angle=0]{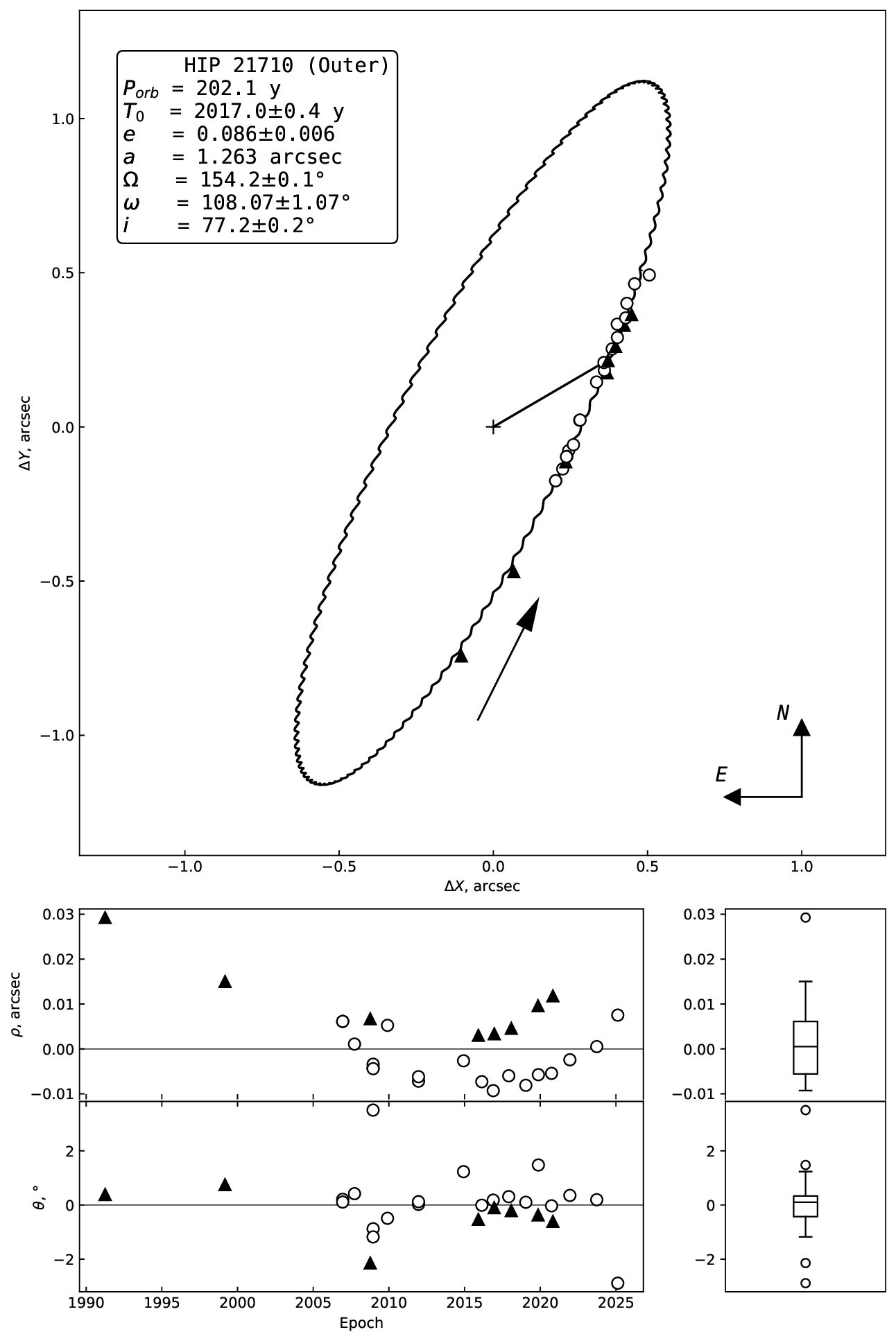}
	\caption{Orbital solutions for HIP~21710. The left panel shows the "first" orbital solution, for which the inner orbit parameters, the wobble factor (f = 0.170), $P_{out}$ and $a_{out}$ were fixed. The right panel shows the "second" orbital solution, for which only the wobble factor (f = 0.170), $P_{out}$ and $a_{out}$ were fixed. All designations are described in the caption of Figure \ref{Fig2}.} 
	\label{Fig4}
\end{figure}

We observed the Aa,Ab pair as part of this study, adding 19 new measurements to the 8 previously published ones (Figure \ref{Fig4}). The orbital solution constructed by \citep{Tokovinin_21_mar} for the tight $Aa_{1}$,$Aa_{2}$ pair of HIP~21710 assumes a value of 46 mas for the semi-major axis. Given such a separation and an eccentricity of 0.426, the system can be resolved at the 6 m telescope. The obtained power spectra for 19 observational epochs suggest that there is no close component in the epochs of 2006.9469, 2007.7264, 2008.9523, 2008.9579, 2009.9081, 2011.9561, 2014.9321, 2016.1393, 2016.8924, 2017.9190, 2019.0469, and 2020.7478, while in the epochs of 2011.9560, 2019.8767, 2021.9574, 2023.7324, and 2025.1221 we observe distortions in the high-frequency region. Unfortunately, we cannot interpret these features of the power spectrum as an image of the third component due to the telescope's aberrations, which are observed on the same scales.

We obtain a change in the separation from 256 mas to 706 mas over the observed range of epochs. With such a change, the accumulation of scale calculation errors may limit the accuracy of our measurements. The discrepancy between our new data and the solution suggested by \citep{Tokovinin_21_mar} may be due to systematic deviations in the separation and position angle.

An outer orbit can be constructed for HIP~21710 since the orbital parameters of the inner one are known. For this purpose, the IDL code orbit3.pro \citep{Tokovinin_17_mar, Tokovinin_17_feb_orb3} was used as described at the beginning of the section. The parameters from \citet{Tokovinin_21_mar} were used as a basis for constructing the orbits, and the parameters of the inner orbit, the wobble factor (f = 0.170), as well as $P_{out}$ and $a_{out}$ were fixed during fitting to preserve the value of the mass sum; this orbital solution is hereinafter referred to as the "first". However, an additional solution was selected that also preserves the value of the mass sum, close to that obtained by \citet{Tokovinin_21_mar}. In that case, the wobble factor (f = 0.170), $P_{out}$ and $a_{out}$ were fixed (hereinafter this orbital solution is called the "second"). Figure \ref{Fig4} and Table \ref{tab2} show both solutions and parameters for the inner and outer orbits. It should be noted that the high accuracy of the obtained orbital parameters is due to the fact that most of them were fixed. The observed section of the orbit is close to linear (Figure \ref{Fig4}), which does not allow us to confidently calculate the orbital parameters of a wide ($\sim 1\farcs6$) pair, small changes in which lead to significant variations in the sought solution for a close pair. Also, as a result of fitting, it is possible to obtain several orbital solutions depending on the parameters that remain fixed and do not change. Note that for large values of the orbital period ($P_{orb}$ $>$ 100 yr) and in the absence of measurements at the moment of orbital rotation (e.g. HIP~28671 from \citet{Mitrofanova_20_jun}), the orbital solution cannot be classified better than grade 4 (“preliminary”). The mutual inclinations are obtained for the "first" and "second" solutions ($\Phi = 47\fdg4 \pm 0\fdg1$ and $\Phi = 122\fdg2 \pm 1\fdg2$, respectively) using equation 6 from \citet{Tokovinin_26_feb}. The mutual inclination calculated using the "first" solution has a value close to $\Phi = 51\degr \pm 10\degr$ from \citet{Tokovinin_21_mar}. The obtained mutual inclination for the "second" solution indicates a retrograde, highly inclined configuration. At the same time, the inclinations of the inner and outer orbits in both solutions have similar values. It can also be concluded that the orbits are not coplanar, which coincides with the conclusion from \citet{Tokovinin_21_mar}. Orbits with mutual inclinations between $\sim$40\degr and 140\degr are subject to Kozai-Lidov cycles, although the eccentricities of the inner $Aa_{1},Aa_{2}$ pair are $e_{in}=0.426$ and $e_{in}=0.414 \pm 0.016$. Further monitoring for HIP~21710(Aa,Ab) is clearly needed to construct a more accurate orbit that would not require fixing the parameters during fitting. It will also clarify the dynamics of the system. Table \ref{tab3} lists the mass sums calculated using the orbital parameters of the "second" solution, the ambiguities of which are due only to the parallax accuracy, because the values of the period and semi-major axis were fixed when constructing the orbit. Taking the value $\mathfrak{M}_{Aa} = 0.9 \mathfrak{M}_{\odot}$ from \citep{Tokovinin_21_mar}, we obtain $\mathfrak{M}_{Ab} = 0.47 \mathfrak{M}_{\odot}$ for the parallax of Gaia DR3, which is consistent with the mass estimate $\mathfrak{M}_{Ab} = 0.53 \mathfrak{M}_{\odot}$ from \citet{Tokovinin_21_mar}.

{\bf HIP~101181} ($20^{h} 30^{m} 33\fs5, +13\degr 49\arcmin 02\farcs4$; HD~195397) is a binary consisting of an F(6-8)V primary and a G(1-6)V secondary \citep{Horch_11_jun}. The first orbital solution was calculated by \citet{Horch_11_jun}. The system was unresolved in 1998-1999 at 82-inch (2.1 m) Struve reflector of McDonald Observatory \citep{Mason_01_jun}. These epochs do not coincide with the periastron passage, which was also discussed by \citet{Horch_11_jun}.

The orbit of HIP~101181 was slightly improved after adding 22 new measurements to the 7 previously published ones (Figure \ref{Fig3}, right panel). The value of the position angle of the measurement by \citet{Horch_11_jun} was changed by 180$\degr$ during orbit construction. The classification of the orbit as "definitive" (Grade 1) is due to the following factors: almost all phases of the orbital period are covered by measurements with a maximum gap of 0.23, and more than one period has passed since the beginning of monitoring of this system. The system was unresolved in 2013.3309 and 2013.7026 at the 6 m BTA SAO RAS, which is close to the periastron passage. The fundamental parameters are calculated using three parallaxes (Table \ref{tab3}) and agree best with each other when using the Gaia DR3 parallax.

\section{Summary}
\label{sect:summary}

Long-term ground-based monitoring, initiated by the Hipparcos mission \citep{Perryman_97_jul}, and in some cases even earlier (for example, HIP~15633, HIP~19472 and HIP~20751), is undoubtedly of great importance for studying the characteristics of binaries. Speckle interferometric observations at the 6-meter telescope since 2007 have made it possible to significantly increase the number of measurements covering various phases of the orbital motion of the secondaries. Despite the significant number of orbits already constructed, they still need improvement.

The data for HIP~1055 with good observing cadence cover part of the apastron, which allowed us to significantly correct the position of the orbital ellipse. At the same time, the nearest expected epoch of periastron passage will occur only in three decades, which does not allow us to expect significant further improvement of the orbit in the near future. The improved accuracy of the orbital solution of the binary HIP~2532(AB) in the work of \citet{Masda_25_feb} is noteworthy. Apparently, this is due to a small systematic discrepancy in the determination of positional parameters between the data we obtained at the BTA and those previously published by other authors, which were used in the work by \citet{Masda_25_feb}. Note that the closest periastron passage of this system is expected at epoch 2030.2 according to the results of our work or at 2028.755 according to the results of \citet{Masda_25_feb}, which will allow us to resolve the apparent discrepancies in the orbital orientation angles. The points obtained for constructing the apparent ellipse of HIP~15633, as in the case of HIP~1055, cover the apastron of the orbit, but the presence of a small amount of published data for other epochs allows us to find the solution more confidently. The improvement of this orbit is not very significant. HIP~20227 was observed by us in phases near periastron, but the relatively small eccentricity of this system will allow us to improve the orbit even in epochs far from $T_{0}$. The residuals obtained for HIP~20751 and HIP~101181 are small, and the measurements cover different phases of the observed ellipses. Observational data are clearly insufficient to construct an accurate orbit of the triple HIP~21710(Aa,Ab), even though the orbital parameters of the inner $Aa_{1},Aa_{2}$ pair are known with high accuracy. This is facilitated by the large orbital period of the system ($P_{orb}$ $>$ 200 yr) and the fact that the object has been observed only while moving along the linear portion of the orbit.

As part of this study, 6 orbital solutions were improved, and 4 of them were classified as "definitive" (Grade 1). Clearly, the “definitive” orbits of the systems can only be slightly improved in the future. However, additional astrometric measurements at orbital phases close to periastron passage are recommended for HIP~1055, HIP~2532, HIP~15633, and HIP~19472, and further long-term monitoring is clearly required for HIP~20751, HIP~20227, and HIP~21710. Also, when studying binaries, we still encounter objects for which only Hipparcos parallaxes are available (HIP~15633). For such objects, the parallax error dominates the mass determination error. Nevertheless, the mass sums and component masses calculated by the two independent methods are in good agreement when using Gaia parallaxes, as described above and in previous studies \citep{Mitrofanova_24_sep}. When the astrometric measurements obtained by the Gaia mission are published, they will provide a significant advance in our knowledge of the orbital parameters and fundamental properties of the systems included in the long-term ground-based monitoring samples.

\normalem
\begin{acknowledgements}

Observations with the SAO RAS telescopes are supported by the Ministry of Science and Higher Education of the Russian Federation. The renovation of telescope equipment is currently provided within the national project "Science and Universities". The work was performed as part of the SAO RAS government contract approved by the Ministry of Science and Higher Education of the Russian Federation.

This work has made use of data from the European Space Agency (ESA) mission {\it Gaia} (\url{https://www.cosmos.esa.int/gaia}), processed by the {\it Gaia} Data Processing and Analysis Consortium (DPAC, \url{https://www.cosmos.esa.int/web/gaia/dpac/consortium}). Funding for the DPAC has been provided by national institutions, in particular the institutions participating in the {\it Gaia} Multilateral Agreement.
This research has made use of the SIMBAD database, operated at CDS, Strasbourg, France.

\end{acknowledgements}
  
\bibliographystyle{raa}
\bibliography{art_2026_bib}

\end{document}